\documentstyle[12pt]{article}
\begin{document}
\begin{center}
{\Large Canonical measure and the flatness of a FRW universe}\\
\bigskip
{\bf D.H. Coule}
\bigskip
\\
Dept. of Applied Mathematics,
University of Cape Town\\
Rondebosch 7700, South Africa\\

\begin{abstract}
 We consider the claim of Hawking and Page that the canonical measure
applied to
Friedmann-Robertson-Walker models with a massive scalar field can solve
the flatness problem i.e. $\Omega\simeq1$, regardless of Inflation
occurring or not. We point out a number of ways this prediction, which
relies predominantly on  post-Planckian regions of the classical
phase space, could break down.\\
 By considering a general potential
$V(\phi)$ we are able to understand how the  ambiguity
for $\Omega$ found by Page in
 the $R^2$ theory, is present in general for scalar field models
when the potential is bounded from above. We suggest reasons why
such potentials are more realistic, which then results in the value
of $\Omega$ being  arbitrary.

 Although the canonical
measure gives an ambiguity (due to the infinite measure over arbitrary
scale factors) for the possibility of inflation,  the inclusion of
an  input from quantum cosmology could resolve this ambiguity.
 This could simply be that due to a ``quantum event''
the universe started small; and provided
a suitable scalar potential is present an inflationary period
could then be ``near certain'' to proceed in order to set $\Omega$
infinitesimally close to unity. We contrast the measure obtained in
this way with the more usual ones obtained in quantum cosmology:
the Hartle-Hawking and Tunneling ones.

\end{abstract}
\end{center}
\newpage
{\bf 1 Introduction}

One of the reasons for the introduction of inflationary universe
models was to resolve the flatness problem-(see eg. refs.[1,2] for
a review). Namely that the universe is
very close to its critical density $\rho_{cr}$ i.e $\Omega\equiv
\rho/\rho_{cr}$ is unity.
There is some doubt that inflation can
entirely resolve this problem [3-5]: if $\Omega$ is initially far from
unity then either recollapse or insufficient inflation would
instead result.

It is also the case  that at present most of the experimental
evidence points to an open universe $0<\Omega\leq1$, see eg. ref.[5]
for a review of this topic. It is therefore of importance to
understand what values of $\Omega$ are to be expected in FRW universes.
This is where the question of measure comes in: what is a typical
solution of the system of equations describing the universe, and what
does it imply for parameters (eg. $\Omega$)
 that we measure experimentally ?\\

{\bf 1.1 The flatness question}

Let us first consider how such a flatness
problem occurs in classical cosmology.
To see this consider a FRW universe with
a perfect fluid equation of state: $p=(\gamma-1)\rho$,  p=pressure.
The Hamiltonian equation is given by
\begin{equation}
H^2+\frac{k}{a^2}=\rho
\end{equation}
with $H$ the Hubble parameter and $k$ the spatial curvature (we use
Planck units throughout and set $8\pi G/3\equiv 1$). There
is also  the continuity equation
\begin{equation}
\dot{\rho}+3H\gamma\rho=0 \;\;\; \Rightarrow \rho =\frac{\rho_o}
{a^{3\gamma}}\;\;\; (\rho_o={\rm constant})
\end{equation}
 Since $\Omega=\rho/H^2$ we can write
\begin{equation}
\Omega=\frac{\rho}{\rho-k/a^2}=\frac{\rho_o}{\rho_0-ka^{3\gamma-2}}
\;\;\;.
\end{equation}
If the strong energy condition is satisfied i.e.
$\gamma>2/3$ then as
the scale factor $a\rightarrow 0$ , $\Omega$ is set initially to
one.
 Whereas if $\gamma<2/3$ then as $a\rightarrow 0$, $\Omega$ is
sent towards $\infty$ for $k=1$, and zero for $k=-1$.\\

For time $t$ increasing to the future,
the value of $\Omega$ diverges away from 1 whenever
the strong energy
condition is satisfied i.e. $\gamma>2/3$. Specifically,
\begin{equation}
\Omega-1\propto t^{2-4/3\gamma}\;.
\end{equation}
 We can therefore
estimate the value of $\Omega$ at the planck time assuming no
inflationary phase occurred and $\gamma\sim 1$ throughout the
evolution of the universe. The age of the universe today is
$\sim 10^{90}t_{pl}$
 \footnote {Assuming the universe is described by a dust model
$\gamma=1$,  the actual age is more like
$\sim H^{-1}\simeq 10^{60}t_{pl}$ but this still results in a bound
$\Omega=1\pm 10^{-40}$. The size of the present universe is at
least $10^{60}L_{pl}$.}
where $t_{pl}$ is the planck time. Using expression (4), we can
relate $\Omega$ at different times as:
\begin{equation}
\frac{(\Omega-1)_{now}}{(\Omega-1)_{then}}\approx
\frac{(10^{90})^{2/3}}{1}\approx 10^{60}
\end{equation}
If we assume that today $\Omega\approx 1$ then at the planck time
we require $(\Omega-1)<10^{-60}$
i.e. $\Omega  \sim 1\pm 10^{-60}$ .

In order to achieve this value of $\Omega$ at the Planck epoch $a=1$
requires the constant $\rho_0\geq 10^{60}$, since from  expression (3)
we have $\rho_0=ka^{3\gamma-2}\Omega/(\Omega-1)$.

Therefore to explain the flatness problem from a classical standpoint
we need to understand why the energy density or
equivalently the Hubble parameter was so large during the initial
stages of the universe.\\

While still on the subject of FRW universes with perfect fluids we
point out a few quantities of {\em Einstein-DeSitter} (k=0)
models which will later
be useful.
The density $\rho$ is given by
\begin{equation}
\rho=\frac{1}{6\pi\gamma t^2}
\end{equation}
and the scale factor behaves as-see eg.[6]
\begin{equation}
a\simeq \rho_0^{1/3}t^{2/3\gamma}\;\;.
\end{equation}
Although we are primarily interested in scalar field models we will
at times use the perfect fluid as an analogy when it is useful.
\\

{\bf 1.2 Inflation}

We generally assume that given a sufficient amount of inflation the
flatness problem can be solved. This is provided the initial conditions
are not, in some sense, too extreme. This is because you can
always extrapolate back any value of
$\Omega$ you choose after a fixed amount of
 inflation, to a corresponding  initial value [3]. Therefore
whether inflation can solve the flatness problem
depends on the measure for initial conditions. If a set amount of
inflation say $\sim 100$ e-foldings does not resolve the
flatness question we will call such measures extreme. The measures considered
in this paper: the canonical and those from quantum cosmology, are
generally found not to be extreme . If they  undergo a certain amount of
inflation they should  resolve the flatness question: i.e. set $\Omega$
infinitesimally close to unity.

The question we mostly address is whether inflation occurs or not? This
depends on the form of the scalar potential $V(\phi)$ and so only a
subset of possible potential allow inflation. For this reason we first
try to resolve the flatness question without requiring inflation {em
per se}. Even if the flatness problem is resolved without
resorting to inflation there are still a
 number of
other reasons why it might be required: for example, to resolve  the horizon
question [1,2]. It is therefore important to know how likely inflation is
 to happen, for a given form of the potential $V(\phi)$.

By the way, inflation, which has $\gamma\approx 0$ would
 lead to a problem if continued back to $a=0$  and indeed
defeat its own purpose if it initially has $\Omega>>1$ or $\Omega\sim 0$.
 Inflation is however not usually assumed to occur immediately from
the singularity but at $a\sim 10^{-30} cm $ [1,2]. \\

{\bf 2 The measure for $\Omega$}

While investigating inflation people have tried to put a measure
on inflationary solutions compared to non-inflationary ones in order
to give probabilities for various quantities such as $\Omega$ in
a typical universe [7-9].  These studies have usually considered equipartition
of initial conditions at an initial planck energy density, when
the classical equations are first expected to be valid. By varying the
initial spatial curvature and Hubble parameter in this way it is found
 that inflation is highly probable [7], with the strong proviso that an
inflationary potential has already been chosen: for example a massive scalar
field $V(\phi)=1/2m^2\phi^2$ with $m<<1$. A
different approach, related to equipartition,
 is found in
ref.[8], while recently a measure using the DeWitt metric has been found [9].

Perhaps a more rigorous measure,
introduced by Gibbons, Hawking and Stewart (GHS) [10],
 is that using a symplectic 2-form $\omega$
which
describes an invariant volume in the phase space (of
dimension 2n) of solutions \footnote {The measure is actually
the symplectic form
to the power $n-1$ , but we only consider models with $n=2$.} . This is
possible because the models under investigation are
even dimensional Hamiltonian systems.  For such models
the Lie derivative of $\omega$ with
respect to the Hamiltonian vector field $X$ is zero: $L_X\omega=0$. There
is some doubt that this would hold for a more general relativistic system
but for the simplified models with few degrees of freedom it would seem
appropriate [11].

Within this approach we
 consider the claim of Hawking and Page [12] that the GHS
canonical measure predicts that the universe is spatially flat
i.e. that the density parameter $\Omega=1$, regardless of Inflation
occurring or not. For the perfect fluid model this would correspond to the
prediction that $\rho_0\rightarrow \infty$, but note that
this also implies $a$ and
$\dot{a} \rightarrow \infty$ cf. eq.(7). We later comment on the
validity of this result and find that it implies vastly post-Planckian
scales.\\

We will essentially use the same equations and notations as presented
in ref.[12]. This considers a massive scalar field model described by the
following equations
\begin{equation}
\dot{a}^2=a^2\dot{\phi}^2+m^2a^2\phi^2-k
\end{equation}
\begin{equation}
\ddot{a}=-2a\dot{\phi}^2+m^2a\phi^2
\end{equation}
\begin{equation}
\ddot{\phi}=-3\frac{\dot{a}}{a}\dot{\phi}-m^2\phi
\end{equation}
It is helpful to define a rescaled scale factor $\alpha$ such that
$\alpha=\ln a$. Using the dimensionless variables
\begin{equation}
x=\phi \;\;\; y=\frac{\dot{\phi}}{m} \;\;\; z=\frac{\dot{\alpha}}{m}
\;\;\; \eta =mt
\end{equation}
the system of equations can be written in the form [12]
\begin{equation}
\frac{dx}{d\eta}=y
\end{equation}
\begin{equation}
\frac{dy}{d\eta}=-x-3yz
\end{equation}
\begin{equation}
\frac{dz}{d\eta}=x^2-2y^2-z^2
\end{equation}
plus a constraint
\begin{equation}
\frac{k}{m^2a^2}=x^2+y^2-z^2
\end{equation}
$k$ is the spatial curvature $\pm 1,0$.
This model or in fact that with any scalar field potential $V(\phi)$
is characterized
by an initial  stiff regime as $a\rightarrow 0$,
 essentially because the kinetic
term $\dot{\phi}^2\sim a^{-6}$ dominates as the singularity is
approached. It therefore behaves like a perfect fluid with $\gamma
\simeq 2$.

For such models, the canonical measure is given by [12]
\begin{equation}
\omega=-d(a\dot{a})\wedge da +d(a^3\dot{\phi})\wedge d\phi\;\;.
\end{equation}
In this model there are 4 phase-space variables ($a, \dot{a}, \phi,
\dot{\phi}$), plus one constraint, making 3 independent variables.
One of these is the parameter along the trajectory, i.e. the ``time'',
so there is a remaining 2 parameters describing the model. The measure is
independent of the choice of 2 dimensional initial data surface chosen
provided each solution hits this surface only once [10,12]. This might not
be true for $k=1$ in that some solutions recollapse before a given
scale factor or energy density is reached - but we ignore this
possible complication.

If we consider the measure on an initial data hypersurface
$a=a_0=constant$ then this simplifies to [12]
\begin{equation}
\omega=a_0^3d\dot{\phi}\wedge d\phi \;\;.
\end{equation}
Using the notation of ref.[12] such that  $\theta= \tan^{-1} (x/y)$
this can be written in the form \footnote {Strictly speaking this
expression does not treat the spatial curvature curvature correctly,
but this only
cause a difference if $z$ or $\rho$ had restricted ranges, see
later eq. (23).}
\begin{equation}
\omega=ma_0^3zdz\wedge d\theta=m^{-1}a_0^3d\rho\wedge d\theta\;\;.
\end{equation}
where the energy density is
$\rho\equiv\dot{\phi}^2/2
+m^2\phi^2$.
The measure, or rather its integral,
 therefore diverges as the Hubble parameter $z\equiv
H\equiv m^{-1}\dot{\alpha}$ or the energy density $\rho$
 is taken to infinity. This occurs because the
variables $z$ or $\rho$ are allowed to have unrestricted ranges, while
still corresponding to regions of the classical phase space.

This divergence in the measure as $\rho \rightarrow \infty$ is the
reason for the prediction that $\Omega=1$. To see this another way,
note that
the Hubble parameter is related to $\Omega$ in the following way
for $k\neq 0$ [3,4]
\begin{equation}
H^2=\frac{k}{a^2(\Omega-1)}
\end{equation}
Only for $H$ or $\rho \rightarrow \infty $ do we have $\Omega=1$ at
fixed $a$.
 It is
important to realize that there is only one solution in the
measure with exactly $ k=0$ or $\Omega=1 $. The solutions are
instead infinitesimally close to $\Omega=1$ as $H\rightarrow \infty$.
Rewriting the measure in terms of $\Omega$
\begin{equation}
\omega=m^{-1}a_0\frac{k}{| \Omega-1| ^2} d\Omega \wedge
d\theta
\end{equation}
we see the divergence as $\Omega\rightarrow 1$.

In Fig.(1) we sketch the probability distribution $P(\Omega)
\equiv \omega $ for $\Omega$
which is taken from this expression (20). This shows that
$\Omega$ is peaked around one regardless of inflation since there
is no need to insist on $m<<1$ to obtain this result. It remains
even for potentials too steep $m\geq1$ to have inflationary  behaviour.

 The divergence
occurs as $\Omega\rightarrow 1$ but actually $\Omega=1\pm\epsilon$ and
$\epsilon>\rho^{-1}_{max}$. This argument is got from substituting
$H^2=\rho/\Omega$ into eq.(19).
 If there are limits on $\rho$ then $\epsilon\neq
0$ and $\Omega$ is not exactly one (except for the one solution where k=0).

Also if there are any upper limit of the energy density $\rho$
i.e. $0<\rho\leq \rho_{max}$
or equivalently the Hubble parameter $0<z\leq z_{max}$ then the measure
is instead finite and the flatness of the universe is not necessarily
assured. Let us estimate what this upper limit should be in order to
predict that $\Omega$ be one today.

The massive scalar field model,
  after any possible inflationary phase,
 oscillates around the minimum
of the potential with an equation of state
like a dust model $\gamma=1$ [12]. If we assume that no inflationary phase
occurred, then as found in section (1.1),
we require $\Omega\sim 1\pm10^{-60}$ at the Planck epoch.
{}From the expression (19) for $ H$ this corresponds to an Hubble parameter
$H^2\geq 10^{60}M_{pl}^2$, where we would expect quantum gravity effects
to dominate when $H^2\sim M_{pl}^2$ !
We can estimate the upper bound on $H$  over which the integral of the
measure, proportional to $H^2$ should
be taken in order that today the probability be $99\%$ that $\Omega\simeq 1$
. A simple consideration of areas $\propto H^2$ gives that the measure needs
to be valid up to $\sim 10^{61} M_{pl}^2$ in order to predict $\Omega\simeq
1$ .
If we can assume that $H$ is in the range $0<H<\infty$ then the measure
$\propto H^2$ is dominated by extremely large $H\rightarrow \infty$.
Because $\Omega = (H^2+k/a^2)/H^2$  and $H^2>>k/a^2$ , $\Omega$ is pushed
extremely close to 1. Any subsequent phase which satisfies the
strong energy condition ($\gamma>2/3)$ would take an extremely long
time before $\Omega$ diverges from 1.
 If however we have an upper bound on $H=H_{u}$ above which the
theory is no longer valid and $H_u<10^{30}M_{pl}$ then today $\Omega\simeq
1$ would not be expected.\\

{\bf 2.1 Reasons for $\Omega\neq 1$} \\
a) Limits on the energy density $\rho$

We have seen that the measure $\propto H^2$ or $\rho$ requires an
unrestricted range in order that the flatness   of the universe
be resolved without resorting to the subset of cases that have
inflation. Note however that although we require $\rho \geq 10^{60}$
at the Planck epoch,
we do not require it to be infinite otherwise it would not appear
consistent with the density of the universe today (modulo
uncertainties in the age and size of the universe, plus expansion
rates).

Because  $\Omega$  is not exactly one it will diverge away from 1
for arbitrary large scale factors. For $k=1$ universes $\Omega\rightarrow
\infty$ before then recollapsing, while
 for $k=-1$ universes $\Omega\rightarrow 0$. \\
 Since
\begin{equation}
\Omega=1+\frac{k}{a^2H^2}=1+\frac{k}{\dot{a}^2}
\end{equation}
the measure at fixed $\Omega$
is equivalent to the measure at fixed $\dot{a}$. In ref.[12] the measure
for $\dot{a}=0$ or $\Omega=\infty$ was obtained and found to be finite,
in contrast to the infinite total measure of solutions.
This calculation only differs by some overall constant if the measure
is done at some other
choice of $\dot{a}$ or $\Omega$, essentially because eq. (15)
has the form
\begin{equation}
x^2+y^2=\frac{\dot{a}^2}{m^2a^2}+\frac{k}{m^2a^2}=\frac{const.}{a^2}\;
\;\; {\rm at\; fixed} \;\;\dot{a}\;\;.
\end{equation}
Note that the numerics of the calculation  for $\dot{a}=0, k=1$ ($\Omega=
\infty$)
turn out the same  for $\dot{a}
=\sqrt{2}, k=-1$ ($\Omega=1/2$).
 If however $\rho$
is restricted the total measure is finite and any value of $\Omega$ is
roughly equally probable for arbitrary scale factors $a$. This statement
could be made more precise but requires the initial upper bound for $\rho$ and
at what scale factor you  measure $\Omega$ to be known. The important point
however is, the claim that $\Omega=1$ for arbitrarily large $a$ requires
the measure be dominated by $\rho_{max}\rightarrow \infty$. Or to see this
another way: the measure eq.(18) when the curvature is included is modified
by letting $z\rightarrow z\ast$ , where
\begin{equation}
z\ast^2=\left (z^2+\frac{k}{m^2a^2}\right ).
\end{equation}
If $z\leq z_{max}<\infty $
we can always reduce the scale factor so that the curvature
dominates i.e $ |k|/m^2a^2> z^2$ and so set $\Omega\neq 1$.

It has been pointed out to me [13] that the problem of requiring vastly
post-Planck-epoch densities can be avoided by considering the measure
not at fixed scale factor but rather at a fixed energy density (which can
be independent of the scale factor as the scalar curvature becomes
negligible [13] cf. eq.(6)). This
is claimed to be more physically reasonable since we have no experimental
evidence for an upper limit on $a$ but we do know the present energy density
within certain limits.

Let us investigate this by considering the canonical measure
for fixed energy density $\rho$ given by the expression [12]
\begin{equation}
\omega=-\dot{\phi}d\phi\wedge d(a^3)=2m^{-1}\rho \sin^2\theta d\theta
\wedge d(a^3)\;\;.
\end{equation}

This is dominated by the divergence at large scale factors $a$
\footnote{We ignore doubts that this seems contradictory in that if the
initial curvature $k=1$ dominates $a$ cannot become large.}
and this aspect was used in ref.[12] to suggest that the curvature $k/a^2$
is correspondingly small at fixed $\rho$ so giving the prediction that
$\Omega=1$. Does this not mean that we could
consider only energies below the Planck epoch and still obtain the result
that the measure predicts that the universe is flat?. The problem is that
the solutions start initially at $a=0$ [12] and then by the Planck time the
scale factor has already tended to infinity. This means that $\dot{a}$ is
now the quantity which is massive in units of $M_{pl}^2$ , which in turn
sets $\Omega $ extremely close to unity cf. eq.(21). But such
extreme values of ``velocity''
$\dot{a}$ beyond the Planck size  $\sim 1$ will induce quantum effects
eg. particle creation, and so actually requiring $\dot{a}\geq 10^{30}
M_{pl}^2$ is extremely suspect. It might also cause problems for
nucleosynthesis if the expansion rate is too fast. Any upper
limit in $\dot{a}$ below this value again sets the total
measure (in eq. (24) )
finite.

  Because of this the value of $\Omega$ would no longer be expected to
be 1 in the universe today, at time $10^{60}$ Planck units later.

The problem of requiring the classical equations to be
valid in extremely quantum gravity regions in order to claim $\Omega=1$
is therefore still
present when we consider the measure
at fixed energy density, and is rather a consequence of the fact
that the phase plane
trajectories of the classical equations
are all appearing from the initial singularity, which is
surrounded by a quantum gravity region through which the solutions
have to pass. \\

There is another reason for suggesting a limit $\rho\leq \rho_{max}$
: to be consistent with a universe described by a finite value
of $a$.\footnote{ There is an
element of philosophical prejudice in this requirement.}
It is unclear how the predictions from the measure, which results in
the claim that $\rho\rightarrow \infty$ or $a\rightarrow \infty$ could
fit the description of our universe. \footnote {Note another way of seeing how
the claim that $a\rightarrow \infty$ comes about:  from eq. (19)
we obtain $k/a^2=H^2(\Omega-1)$
so we see that as $\Omega\rightarrow 1$ then $a\rightarrow \infty$.}
 With the proviso that future experimental
evidence (coming especially from a better knowledge of the deceleration
parameter $q$) will in future give an finite upper limit to the scale factor
i.e. $q\neq 1/2\Rightarrow a\neq\infty$. Only when $k=0$ or $q=1/2$
does a finite age of the universe not imply a finite scale factor.
Recall also that the constant
$\rho_0$ for the perfect fluid model is related to $q$ as:
\begin{equation}
\rho_0=\frac{2q}{H(2q-1)^{3/2}}\;.
\end{equation}
So $q=1/2$ sets $\rho_0\rightarrow \infty$.

 The big bang model with $k=1$,
starts out with a small scale factor and expands to a possible
large value but staying finite throughout its evolution. Contrast
this with the measure at fixed scale factor $a$ which suggests
$\rho\rightarrow \infty$, or at fixed energy density that $a\rightarrow
\infty$. Although some
upper limit on $\rho$ or $a$ (maximum possible scale factor)
 consistent with our universe
could be larger that that required to solve the flatness problem it
seems clear that an upper limit over which the measure can act
must exist.

 This would mean  that at some stage
the measures cannot be applied over unrestricted  values of the parameters
$\rho$ or
$a$ so giving a finite measure for
such models.   We would be left with the problem of why these restrictions on
$\rho$ in order to provide a realistic universe model are consistent with large
enough values to also solve the flatness problem. In other words we
require the value of $\rho$ at the Planck scale to be finite, say
less than $\rho_{max}$ but
greater that $\rho_{flat}\sim 10^{60}$ Planck units in order to
solve the flatness problem. According to the measure
we should expect $\rho\rightarrow \infty$ but rather we require
the prediction $10^{60}<\rho\leq\rho_{max}$.
\\

In summary of this section, the GHS measure $\propto \rho$ can indeed
explain the flatness of the universe but assumes the theory to be
valid up to enormous energy densities or velocities.
 If the theory becomes invalid
before this value is achieved then the prediction of $\Omega=1$
cannot be sustained without resorting to
mechanisms like inflation. Because only a certain subset of possible
potentials can have inflation this is a more restrictive requirement.

 To be consistent with our universe the
measure must break down at some finite energy density $\rho_{max}$
but this could be at a value which still allows the explanation
of $\Omega=1$.
\\

b) Limits on the scalar potential $0\leq V(\phi)\leq V(\phi)_{max}$.\\

 There is an alternative way in which the the universe need
not be spatially flat even assuming $\rho\rightarrow \infty$.
We consider the case of an arbitrary potential $ V(\phi)$
but with unrestricted $\rho$. In analogy with the angle $\theta$ which
was introduced to parameterize the massive scalar field case, we can
introduce the angle $\Psi$, such that cf. ref.[14]
\begin{equation}
\dot{\phi}=\rho^{1/2}sin(\Psi)
\end{equation}
\begin{equation}
V(\phi)=\rho \cos^2(\Psi) \;\;.
\end{equation}
 Define $f(\phi)=(V(\phi)/\rho)^{1/2}$
so that $\Psi=cos^{-1}f$
Then
\begin{equation}
\frac{d\phi}{d\Psi}=-\frac{\sqrt{1-f^2}}{f'}\;\;\; ('\equiv df/d\phi)
\;\;.
\end{equation}
Substituting for $d\phi$ into the measure for fixed
energy density eq.(24), gives,
\begin{equation}
\omega=-\dot{\phi}\frac{\sqrt{(1-f^2)}}{f'}d(\Psi)\wedge d(a^3)
\;\;.
\end{equation}
Or in terms of $V(\phi)$ we obtain
\begin{equation}
\omega = \frac{\sqrt{V(\phi)}}{V'(\phi)}\rho \sin^2\Psi d(\Psi)\wedge d(a^3)
\end{equation}
For this arbitrary potential the measure still is dominated by the $a^3$
divergence, but in contrast
 to the massive scalar field case (see Fig. (2))  we can now
obtain an infinite measure even for finite $a$ when
1) $ df/d\phi=0$ or equivalently $dV/d\phi=0$
and
2) $1>f$ which implies $\rho>V(\phi)$.
\\

When calculating the measure at fixed $\Omega$ one obtains expressions
of the form (cf. eq.(4.12) in ref. [12]),
\begin{equation}
\omega=\frac{const. \sqrt{V(\phi)}}{V'(\phi)}\left(1-3sin^2\Psi
\right )da\wedge d\Psi
\end{equation}
where the constant is related to that on the RHS of eq.(22).
 When integrated over allowed values of $\Psi$ this gives a total measure
for scale factor $a$
\begin{equation}
\mu \sim \pi\frac{\sqrt{V(\phi}}{V'(\phi)}a\;\;.
\end{equation}
Because of the factor $V'$ in the denominator there is a
possible infinite measure for any  fixed $\Omega$ at a given scale
factor $a$. Any value \footnote{
Provided the energy density is large enough to allow the $V'$
region to be reached, so  $\Omega$ is above a small lower bound .}
of $\Omega$ would have roughly the same
probability of occurring and without some extra mechanism to
restrict it would appear arbitrary.\\

 For $k=-1$ the universe will
expand until possibly moving onto the  plateau where $V'(\phi)=0$.
Once there an
inflationary epoch could result and so ensure in this case that
$\Omega=1$ is reached to sufficient precision. An example of this would be a
 potential of the form $V(\phi)
=tanh^2\phi$, see Fig.(3)\\

However there is an ambiguity in the GHS measure for
inflation to occur [12], essentially because of the $a^3$ divergence in the
measure which prevents you concluding that even ``inflationary looking''
potentials must inflate. We suggest later how this problem might be
resolved by appealing to quantum initial conditions.
 Other bounded potentials with non-inflationary plateaus such
as $V(\phi)=cos^2\phi$ would also not allow the ambiguity of $\Omega$ even for
$k=-1$ universes, to be resolved, see Fig.(4).
\\
In general any potential which tends to a constant say $V_1$
(i.e bounded) and the
energy density $\rho$ is greater than$ V_1$ would not resolve the
flatness of the universe.
 This enables us to understand the $R^2$ model considered
in ref.[14], where it was not apparent whether the non-resolution of the
flatness was caused by some aspect of the $R^2$ model. We now see
that this non-resolution persists for ordinary gravity together with a
suitable scalar field. \\

 The $R^2$ model is conformally related to a scalar theory
with potential
\begin{equation}
V(\phi)=\frac{1}{2\epsilon}\large \left (1-\exp(-2\phi)
\large \right )^2
\end{equation}
where $\phi=1/2\ln(1+2\epsilon R)$ and we limit $\phi\geq0$.
In this case
$f'=exp(-2\phi)$ which  tends to
zero as $\phi \rightarrow \infty$ - so giving the infinite
measure.\\

This fall off in the potential towards a constant is a fairly common
feature as theories are pushed to their breaking points eg. the over
driven
pendulum becoming circular motion. It might
therefore be the case that as the quantum gravity regime is
reached the
potential reaches a plateau. See the related discussion in ref.[15]
that due to the conformal anomaly the potential is bounded.

Another example of the flatness being unresolved is
that of a cosmological constant
and a massless scalar field (axion). If such quantities are present
in the universe
then the classical measure would not unambiguously give $\Omega=1$.
 Other degrees of freedom cf. the Bianchi(1) model with its extra
unrestricted variable (see eq.(3.7) in ref. [16]),
 have a similar effect to the bounded potential
in that another variable has an infinite
range which introduces ambiguities in predictions from the measure.
If the measure was extended to inhomogeneous models we would expect
such problems to remain, and the bounded potential can be taken as
simulating the problems inherent when  an anisotropy or inhomogeneity
is present.

 It has been suggested [13] that this bounded
potential still predicts $\Omega=1$. By restricting $\rho=\rho_{anth}
<V_1$ the singularity at $V'=0$ is avoided,
where $\rho_{anth}$ is the maximum density from anthropic grounds
consistent with life.\\
 However if we are assuming a classical description
since the initial singularity, we are primarily interested in the measure
applied to the relevant physics applying to the earliest stages
of the universe. The evolution (trajectory of the universe)
is set during this initial high density epoch and a
subsequent change in the physics (at lower $\rho$)
 with its different measure will not
change this. The measure does not determine dynamics but only allows you
to give probabilities amongst various solutions allowed by the
dynamics, because in the early universe with its possible $R^2$ theory
and large energy densities ($\rho \geq V_1$) this ambiguity in the flatness
is possible. Even if later in the evolution of the universe such theories
are no longer valid, the ``the die has been cast''and we should not
attempt to predicts things
using re-calculated measures. For example, we could obtain erroneous
conclusions
that the universe goes from non-flat to flat suddenly as a massless axion
acquires a mass after some QCD phase transition.

 So far we have been interested in showing reasons why $\Omega$ might
not be unity within the canonical measure framework regardless of inflation
. Indeed the scheme
works for any non-bounded potential and so is more general than inflation
which requires a ``flat'' potential.\\

If the measure acts over an unlimited range of initial
energy density then $\Omega$ is set infinitesimally close to
one and would still be $1$ at a time $10^{60}t_{pl}$ later. This occurs because
the massive scalar field behaves initially like that of a
perfect fluid with the strong energy condition satisfied, actually $\gamma
\sim 2$. The analogy with the perfect fluid model is that at the initial
start of the Big Bang
we are choosing at random the constant $\rho_0$ which is uniform within
the range $0\leq\rho_0\leq \infty$. A typical universe will therefore
have $\rho_0\rightarrow \infty$, see also ref.[5].   In this regard the
canonical measure simply puts this argument (of choosing random
constants) on a more formal footing for
the  scalar field case.\\
 The energy densities (or values of $
\dot{a} $) involved in dominating
the
measure are enormous and the classical theory will surely have been
superceded by a quantum gravity regime  at the relatively low energy
scale $\rho \sim$ Planck size.\\

{\bf 3 Quantum Era}\\
Let us consider briefly the values of $\Omega$ that might occur if the
universe was initially created in a quantum event with an Hubble parameter
$H^2\sim 1$ or energy density $0<\rho < 2$ (for an initial size $a \sim 1$).
We are assuming that the
classical equations are valid immediately the universe is created.
 From expression (17) notice that $\Omega$ for both closed and
open universes diverges away from one for  scale factors approaching one
from above, (i.e as $a
\rightarrow 1$). This  means that typically $\Omega$ is in the range $0<
\Omega\leq2$. For larger initial scale factors $\Omega\rightarrow 1$
as $a>>1$ . In order to obtain $\Omega>>1$ we would require
initially $H^2\simeq 0$ .\\

Inflation
can help a much wider spread of initial values of
$\Omega$ become spatially flat. In the previous example of an initial
quantum event with $H\sim 1$, only when $\Omega \simeq
0$ or $\rho \simeq 0$ might we fail to achieve $\Omega=1$
 due to  insufficient matter being present to give inflation.
It is therefore important to know from a possible quantum gravity
theory the possible value of the initial Hubble parameter.

 One
approach is that of the Wheeler DeWitt equation which gives a
solution for the wave function of the universe $\Psi$. For a
simple $k=1$ model of a constant scalar field $V(\phi)=const.$ the
probability of the initial Hubble parameter H is given by
\begin{equation}
\Psi_{H}^2\sim \exp\left (\pm 1/H^2\right )
\end{equation}
where the + and - signs correspond to the Hartle-Hawking and
Tunneling boundary conditions respectively see eg. [17,18]. The Hartle-Hawking
case would suggest that $H\simeq 0$ which would suggest an initial $\Omega
>>1$, and little possibility of inflation to correct this. There is
however some dispute on this interpretation [17,18]. The tunneling
boundary condition on the other hand would suggest a larger initial
Hubble parameter and so a closer to one value for $\Omega$. This could
then be set infinitesimally
close to one by a reasonable period of inflation.

Very little has been done for spatially open models (k=-1) to
see what predictions are made for $H$ and  $\Omega$ and whether a
period of inflation would ensue. Some exact solutions of the WDW equation when
$k=-1$ have been found [19] for conformally coupled scalar fields
, but these are not inflationary models.

Recently a related approach based on a classical change of signature
has been investigated [20]. In the sense that euclidean regions
can be described by the classical equations, but the equations can
in turn be quantized. The WDW equation in this case
 appears only consistent with a tunneling
boundary condition.

A quantum analogue of the GHS measure
which predicts the probability of  $\Omega $
at the start of the classical evolution is required.
In the WDW scheme Gibbons, Grishchuk and Sidorov [21] have made an initial
attempt to define a typical wavefunction, and find that it is
similar to the tunneling one. This usually predicts
inflation provided a suitable potential is
available and so this would suggest  that the result $\Omega=1$ could be
obtained.

A prediction from some measure that $\Omega=1$
without inflation would be preferable, since inflation is
dependent on there being present a suitable potential.

In summary of this section: we would expect from an initial quantum
creation of the universe that $\Omega$ is not too far initially from 1,
depending on boundary conditions: in the language of
section 1.2 we find no evidence of extreme measures.
Such a value of $\Omega$ could be
then
pushed extremely close to 1 by a suitable period of inflation.
Ideally a quantum gravity measure, which had no
need of classical inflation, analogous to the GHS classical one
would be preferable.\\

{\bf 4 Inflationary epoch}

 According to ref.[12]
the canonical measure cannot predict for
certain if an inflationary phase occurs. The measure for the massive
scalar field case
typically has the form [12],
\begin{equation}
\omega\simeq d\phi_1\wedge d(a^3_1)
\end{equation}
where $a_1$ and $\phi_1$ are respectively the scale factor and initial value of
the scalar field when inflation starts. There is an
infinite measure due to the $a^3$ term even for values of the scalar field
below some value that gives sufficient inflation.
 Note that this  assumption that
 the initial scale factor can be anything is the cause of this
ambiguity.  But if we assume from
quantum cosmology that the universe had to start ``small'', then this
ambiguity could be  corrected.\\
This ambiguity is not present in quantum cosmological models since
the initial scale factor is assumed to be of $\sim$ Planck size.
For example the WDW equation gives, in analogy with expression (35)
 a quantum  measure of the form, see eg. [17,18]
\begin{equation}
\omega\simeq \exp\left (\pm\frac{ 1}{V(\phi)}\right )d\phi_1
\end{equation}
where the + and - signs correspond to HH and Tunneling boundary
conditions. This measure is either more peaked around small values
of $\phi$ (HH) or prefers larger values (Tunneling) compared to
the uniform case $\sim d\phi_1$.

 Encouragingly, the measure derived from a classical
change of signature has the form, called $dP^{CS}_V$ in ref. [20],
\begin{equation}
\omega\simeq d\phi_1.
\end{equation}
It is therefore  uniform in $\phi$ as is the classical GHS one: but it
no longer suffers from the divergence over arbitrary scale factors.
 In some sense this approach gives a measure  ``in between''
the purely classical GHS one and those from the usual WDW quantum cosmology
eg. HH.

Although we might have been led to this result ($\omega\simeq d\phi_1$)
by assuming the universe
started ``small'',  it is good to see this result coming from an
entirely different approach and analysis to that of the GHS one.

Once this restriction on the initial scale factor is made the total
measure, at least for unbounded potentials,
 becomes finite and the probability of inflation can be
obtained.  For the massive scalar field model
inflation occurs for certain values of $\theta$: those
corresponding to a large enough scalar field $\phi$. Following
along the lines of ref.(12) the fraction of non-inflationary
solutions $f_{NI}$, for at least $z$ e-foldings of inflation is
\begin{equation}
f_{NI}\sim \frac{m\ln(1+z)}{\rho^{1/2}}
\end{equation}
Provided $m\sim 10^{-4}$ and we require $\sim 100$ e-folding of
inflation then $f_{NI}\sim 10^{-4}$ for energy densities $\rho
\sim $ Planck value. This is a considerable improvement to the
 classical canonical measure reason for setting
$\Omega=1$, which needs to be valid up to energy
scales $\sim 10^{40}$ planck size. We are however still at
energy scales at which we expect quantum gravity effects to
dominate. We also have to understand why the mass $m$ is so
small, if $\rho$ is reduced it needs to become correspondingly smaller
to predict $f_{NI}\sim 0$.   This example could result in the
prediction that $\Omega=1$, if for example the energy density $\rho$ is
limited  and the GHS measure cannot itself give $\Omega=1$. This
is possible because the total measure of solutions is finite and
 makes it possible to give the  fraction of inflationary
solutions among them.\\

 In the case of a bounded potential however there is still the
divergence due to the $V'(\phi)$ term in the denominator.
It still appears not possible to
determine the probability of inflation even for potentials like $V(\phi)
=\tanh^2(\phi) $ which at first sight look very inflationary. There is
still an infinite measure for any value of $\Omega$ and certain of these
 cannot inflate\footnote{This is due mostly to universes recollapsing which
due to the initial data surface we have already potentially underestimated.} .
 However, in the case of $k=-1$
universes it should be possible
once the $a^3$ divergence is eliminated to conclude that inflation is
near certain to occur. This is  provided the initial energy density is above
the plateau  so that as the universe expands it will fall onto the plateau at
random, with an almost certain initial $\phi>\phi_*$, where $\phi_*$ is the
value of the field that gives sufficient inflation $\sim 60$ e-foldings [1,2].
One can see that this scenario is likely in the case of the $R^2$ model
see eq.(6.113) in Ref.[14]. \\
\\
{\bf 5 Anthropic restrictions}

Somehow the anthropic principle should play a role within the measure
question. The measure has been used to define a probability for some
quantity eg. $\Omega$ in a typical universe. This has  depended on the
assumption that  an ensemble of universes is possible. One in effect ``throws
dice'' at the start of the Big Bang
to determine the constant(s) ($\rho_0$ in the perfect fluid case) that
determines all the subsequent evolution of the universe. If the throw is
a ``dud'' (i.e $\rho_0$ small) then the universe recollapses or is empty.
Life will therefore not be around at a time $10^{60}t_{pl}$ later  to question
why such a case happened.

Consider the previously mentioned  case of the
inflationary potential $V(\phi)=tanh^2(\phi)$. We found that
 roughly speaking:  closed universes are near certain to collapse while
open ones are almost certain to inflate. On  anthropic grounds
we could then conclude that inflation {\em did} occur and
that  $\Omega \leq1$ in the subsequent universe
(depending on how much inflation occurred and at what time in the future
we wish to measure $\Omega$).\footnote{ Note that this differs from my
earlier criticism of an anthropic argument where I saw no reason to
try and limit the possible value of $\rho$ in the early universe, only
much later in its evolution do we require $\rho\leq\rho_{anth}$. The
initial high value of $\rho\geq V_1$ was causing an ambiguity in all universes
not just recollapsing ones.}
 In this example there was initially
a  roughly 50/50
chance of either recollapse or inflation occurring. The ``quantum process''
responsible for such a  classical universe would
not have to be repeated  very often in order to get a universe which did
inflate and could later sustain life. Even if we had concluded that
recollapse was vastly more probable, (for the $V(\phi)=tanh^2(\phi)$
potential case) than inflation then on anthropic
grounds we could, if we existed in such a universe, conclude
that inflation did occur anyway. This suggests that the``quantum process''
responsible for the start of such universes could repeatedly
attempt
many tries  and only finally after many
goes did  inflation occur.\\
\\
The eternal inflationary universe model [1,22] tries
to justify this sort of argument:
 why an ensemble of universes is possible. Some of these
universes
will be long lived and so suitable for life to evolve. But this
scenario still requires an initial domain to first start inflating,
which also comes from a singularity.
\\
The singularity theorems
 have recently been extended to include inflationary
regimes, and are still found to
require the existence of singularities even for
open (k=-1) models [23]. \\
The eternal inflationary model
 is therefore not entirely  immune from the measure question. i.e. what if
the possibility of the creation of the first inflating domain
is minute? Would we still have to rely
then on anthropic arguments ? But such worries can anyway wait
until the likely initial
conditions from a quantum gravity theory are better known.
\\
\\
{\bf 6 Conclusions}

We have seen that the canonical measure when all regions of phase
space are allowed predicts $\Omega=1$ for monotonically increasing
scalar potentials. This essentially occurs because the kinetic term
is rapidly diverging to $\infty$ as $a\rightarrow 0$ and so gives
an enormous initial Hubble parameter which sets $\Omega$ initially
very close to one. It would then take an enormous time
(during periods when the strong energy condition is satisfied
$\gamma>2/3$) if it is
to eventually diverge away from unity. It therefore does solve
the flatness problem without appealing to inflation (or even
requiring the presence of an inflationary potential) but at the
cost of assuming the classical equations are valid arbitrarily
close to the
initial singularity.\\
\\
We have given two ways in which this prediction could break down
a) If the maximum energy density (or velocity $\dot{a}$)
over which the measure acts is
less than $\sim 10^{40}$ Planck units or b) there is an upper
limit (plateau) to the scalar potential. The first of these conditions
is especially suspect in light of the massive energy densities ( or
derivatives of the metric $\dot{a}$)
present which are still assumed to obey the classical equations.
We would rather expect the metric to become ``fuzzy'' at around the
Planck scale or quantum processes to invalidate the dynamics.\\
\\
If we assume a quantum event started the universe then a rough
order of magnitude predictions is that $H\sim 1$ and so the
Hubble parameter is not
large enough to
set $\Omega=1$ to sufficient accuracy
. Unless some quantum analogue of the GHS measure predicted
an initial value of $\Omega$ exactly one, we would have to
rely on a subsequent classical mechanism (inflation) to explain
the spatial flatness of the universe.

In the WDW approach, depending on boundary conditions
we might have sufficient inflation to set $\Omega=1$ provided a
suitable mechanism of inflation ( a classical event) is present
after the quantum epoch. This is more restrictive than the GHS
mechanism for the purely classical case, which does
not require an inflationary epoch. Inflation does however
enable the flatness problem to be solved without appealing to
energy scales vastly greater  than the Planck scale.

Even if the flatness problem is resolved, inflation might be
needed to resolve other problems eg. to dilute anisotropy or to
generate perturbations for galaxy formation [1,2]. It is
important to determine the chances of inflation happening.

The classical measure gives an ambiguity for the probability of
inflation occurring (there are $\infty$ solutions of both
inflationary and non-inflationary cases).

This ambiguity can be corrected by means of a simple input from
quantum cosmology: that the initial scale factor  was small.
When this is done the measure ($\omega\sim d\phi_1$) is moderately
inflationary compared to purely quantum cosmological measures.
Moderate because unlike the tunneling case, it is not peaked at large
values of $\phi$, but neither is it peaked, like the HH case, at small
$\phi$. It  also agrees
with the measure found from  a classical signature
change approach.
\newpage
{\bf Acknowledgement}\\
I am very grateful to Don Page for many enlightening  and
interesting e-mail exchanges
during the course of this work.

George Ellis initially drew me to the importance
of the measure question and provided encouragement throughout.

 Also thanks to C. Hellaby, B. Stoeger and
P. Dunsby. \\

 {\bf References}\\
\begin{enumerate}
\item A.D. Linde, {\em Particle Physics and Inflationary Cosmology}
(Harwood, Switzerland, 1990).
\item E.W. Kolb and M.S. Turner, {\em The Early Universe}, (Addison
-Wesley, USA, 1990).
\item G.F.R. Ellis, in {\em  Proceedings of Banff Summer
Research Institute on Gravitation} eds. R. Mann and P. Wesson
(World Scientific, Singapore) 1991\\
 G.F.R Ellis, Class. Quan. Grav. 5 (1988) p.891
\item C. Magnan, Comments on Astrophysics, {In press} 1994.
\item G.F.R. Ellis and P. Coles, {\em The case for open Universes}
in press, Nature, (1994)
\item L.D. Landau and E.M. Lifshitz, {\em The Classical theory
of fields} (Pergamon Press, Oxford) 1975.
\item V.A. Belinsky, L.P. Grishchuk, Y.B. Zeldovich and I.M.
Khalatnikov, Sov. Phys. JEPT 62 (1985) p.195.\\
 V.A. Belinsky and I.M. Khalatnikov, Sov. Phys. JEPT 66 (1987)
p.441.\\
V.A. Belinsky, H. Ishihara, I.M. Khalatnikov and H. Sato, Prog.
Theo. Phys. 79 (1988) p.674.\\
M.S. Madsen and P. Coles, Nucl. Phys. B 298 (1988) p.2757.\\
D.H. Coule and M.S. Madsen, Phys. Lett. B 226 (1989) p.31.
\item H.J. Schmidt, Astron. Nachr. 311 (1990) p.99.
\item H.T. Cho and R. Kantowski, {\em A measure on a subspace of
FRW solutions} preprint (1994)
\item G.W. Gibbons, S.W. Hawking and J.M. Stewart, Nucl. Phys.
B 281 (1987) p.736.
\item M. Henneaux, Nuovo Cim. Lett. 38 (1983) p.609.\\
see also: C. Tzanakis, ibid p. 606.
\item S.W. Hawking and D.N. Page, Nucl. Phys. B 298 (1988) p.789.
\item D.N. Page, private communication
\item D.N. Page, Phys. Rev. D 36 (1987) p.1607.
\item A. Vilenkin, {\em Predictions from quantum cosmology}
preprint (1994).
\item P. Chmielowski and D.N. Page, Phys. Rev. D 38 (1988) p.2392.
\item J.J. Halliwell, in {\em Quantum Cosmology and Baby Universes}
eds. S. Coleman et al. (World Scientific, Singapore, 1991).
\item D.N. Page, in {\em Proceedings of Banff summer research institute
on Gravitation} eds. R. Mann and P. Wesson (World Scientific, Singapore)
1991.
\item D.N. Page, J. Math. Phys. 32 (1991) p.3427.
\item J. Martin, Phys. Rev. D 49 (1994) p.5105.
\item G.W. Gibbons and L.P. Grishchuk, Nucl Phys. B 313 (1989) p.736.\\
L.P. Grischuk and Y.V. Sidorov, Sov. Phys. JEPT 67 (1988) p.1533.
\item A. Linde, D. Linde and A. Mezhlumian, Phys. Rev. D 49 (1994) p.1783.
\item A. Borde and A. Vilenkin, Phys. Rev. Lett. 72 (1994) p.3305.\\
A. Borde, {\em Open and Closed Universes, initial singularities
and Inflation} preprint Tufts University (1994).

\end{enumerate}
\newpage
{\bf Figures}\\

Fig.1 )
The probability of $P(\Omega)$ against $\Omega$. This shows the
divergence in the measure at $\Omega=1$ but in order to approach
exactly 1 ($\epsilon\rightarrow 0$),
would require the energy density $\rho \rightarrow \infty$.
\\
\\
Fig. 2)
A potential which the GHS measure would predict has $\Omega=1$
without requiring inflation. Note
that in this example $V(\phi)=\phi^2$ i.e. $m=1$ and so
is too steep to inflate.
\\
\\
Fig. 3)
An example $V(\phi)=\tanh^2(\phi)$
of a bounded potential which can have any value of $\Omega$
according to the GHS measure. Even though this potential is ``inflationary''
the probability of inflation given by the GHS measure is ambiguous and
so we cannot conclude that inflation must set $\Omega=1$.
With certain
quantum initial conditions which ensured that the evolution started
on the plateau, we could get a ``certain" prediction that inflation occurs
i.e. $\phi>\phi_*$.
\\
\\
Fig. 4)
An example of a bounded potential $V(\phi)=\cos^2(\phi)$
, which also does not have the possibility
of inflating. The potential is too steep to inflate and the scalar field
quickly rolls down to a minimum.
In such cases the value of $\Omega$, at either
a fixed value of the scale factor or the energy density, is arbitrary according
to the GHS classical measure.
\end{document}